\documentclass[sigconf,nonacm]{acmart}
\usepackage{balance}
\usepackage[rightcaption]{sidecap}
\usepackage{xcolor}
\usepackage{latexsym}
\usepackage{tikz}
\usetikzlibrary{positioning,shapes,decorations.pathmorphing}

\usepackage[utf8]{inputenc} 
\usepackage[T1]{fontenc}   
\usepackage{hyperref}       
\usepackage{url}
\usepackage{multirow}
\usepackage{booktabs}       
\usepackage{amsfonts}       
\usepackage{nicefrac}       
\usepackage{microtype}      
\usepackage{xcolor}        
\usepackage{tabularx}
\usepackage{notes-alt} 
\usepackage{multirow}
\usepackage{longtable}
\usepackage{supertabular}
\usepackage{makecell}
\usepackage{amsmath}
\usepackage{graphicx}
\graphicspath{ {imgs/} }
\usepackage{svg}
\usepackage{enumitem}
\usepackage{subcaption}
\usepackage{amsthm, amsmath}
\usepackage[linesnumbered,ruled]{algorithm2e}
\usepackage{xcolor}

\definecolor{blue}{RGB}{17,220,247}
\definecolor{purple}{RGB}{163,115,250}

\newcommand{\up}[1]{\tiny ($\textcolor{green}{\blacktriangle}#1\%)$}
\newcommand{\down}[1]{\tiny ($\textcolor{red}{\blacktriangledown}#1\%)$}

\usepackage{wrapfig}
\newcommand{\mpara}[1]{\medskip\noindent{\bf #1}}

\newcommand{\bert}{\textsc{BERT}}
\newcommand{\bart}{\textsc{BART}}
\newcommand{\gpt}{\textsc{GPT-2}}
\newcommand{\gptt}{\textsc{GPT-3}}
\newcommand{\car}{\textsc{CAR}}
\newcommand{\linear}{\texttt{Linear}}
\newcommand{\attention}{\texttt{Attention}}

\newcommand{\chatgpt}{\textsc{ChatGPT}}
\newcommand{\davinci}{\textsc{Davinci-002}}
\newcommand{\vinci}{\textsc{Davinci-003}}

\newcommand{\babbage}{\textsc{Babbage-001}}
\newcommand{\curie}{\textsc{Curie-001}}
\newcommand{\ada}{\textsc{Ada-001}}
\newcommand{\qd}{\textsc{Query2Doc}}

\newcommand{\ms}{\textsc{MS MARCO}}

\newcommand{\trecdl}{\textsc{TREC-DL}}

\title{The Surprising Effectiveness of Generative Query Rewriting for Training Text Rankers}
\title{Context Aware Query Rewriting for Text Rankers using LLM}

\author{Abhijit Anand}
\email{aanand@l3s.de}
\affiliation{%
  \institution{L3S Research Institute}
  \country{Germany}
}

\author{Venktesh V}
\email{v.Viswanathan-1@tudelft.nl}
\affiliation{%
  \institution{Delft University of Technology}
  \country{Netherlands}
}

\author{Vinay Setty}
\email{vsetty@acm.org}
\affiliation{%
  \institution{University of Stavanger}
  \country{Norway}
}



\author{Avishek Anand}
\email{Avishek.Anand@tudelft.nl}
\affiliation{%
  \institution{Delft University of Technology}
  \country{Netherlands}
}

\begin{document}
\begin{abstract}

Query rewriting refers to an established family of approaches that are applied to underspecified and ambiguous queries to overcome the vocabulary mismatch problem in document ranking.
Queries are typically rewritten during query processing time for better query modelling for the downstream ranker.
With the advent of large-language models (LLMs), there have been initial investigations into using generative approaches to generate pseudo documents to tackle this inherent vocabulary gap. 
In this work, we analyze the utility of LLMs for improved query rewriting for text ranking tasks.

We find that there are two inherent limitations of using LLMs as query re-writers -- concept drift when using only queries as prompts and large inference costs during query processing.
We adopt a simple, yet surprisingly effective, approach called context aware query rewriting (\car{}) to leverage the benefits of LLMs for query understanding.
Firstly, we rewrite ambiguous training queries by context-aware prompting of LLMs, where we use only relevant documents as context.
Unlike existing approaches, we use LLM-based query rewriting only during the training phase.
Eventually, a ranker is fine-tuned on the rewritten queries instead of the original queries during training.
In our extensive experiments, we find that fine-tuning a ranker using re-written queries offers a significant improvement of up to \textbf{33\%} on the passage ranking task and up to \textbf{28\%} on the document ranking task when compared to the baseline performance of using original queries.

\end{abstract}
\maketitle
\section{Introduction}
\label{sec:intro}

The \textit{vocabulary mismatch} between user queries and the documents is a well known problem in the field of Information Retrieval (IR). 
The user information needs usually represented in the form of keyword queries can be ambiguous and may not be lexically similar to the documents. 
The queries could be under-specified, rendering it difficult to understand the user intent, which affects the downstream retrieval performance. 
For instance, the query \texttt{define sri} could refer to ``sanskrit word sri'' or ``\textbf{S}ocially \textbf{R}esponsible \textbf{I}nvestment''.
To disambiguate the users' information need, many approaches have been proposed to address the vocabulary gap.

Classical approaches like pseudo-relevance feedback mechanism \cite{lv2009comparative, ir_query_exp} expand the original query with keywords from top-ranked results to the original query. Alternate term-based approaches represent the queries in a \textit{continuous vector space}, followed by a KNN-search to expand query terms \cite{grbovic2015context,zamani:2017:relevance-based-word-embedding}. There exist other term-based approaches that use seq2seq model~\cite{he2016learning} to generate query rewrites. 
However, all the above approaches are term-based approaches that improve retrieval but not necessary in document ranking. 

Going beyond keywords, researchers have explored the possibility of natural language question generation for clarifying queries~\cite{rao2018learning} or rephrasing original queries using deep generative models for alternative formulations of the original query~\cite{zerveas2020brown}.
More recently, document expansion approaches have been adopted to improve retrieval performance. \textsc{Doc2Query}~\cite{nogueira2019document} employs neural models to predict queries relevant to documents and enhances document representations. 
Especially with the advent of large language models or LLMs~\cite{gpt3_prompting,gpt3incontext}, recent approaches have started to investigate the utility of LLMs to aid in enhancing query and document representations~\cite{wang2023query2doc,gospodinov2023doc2query}.
LLMs are promising because encode world knowledge \cite{yu2023generate} from pre-training can be employed to reformulate the under-specified queries. 
In principle, the LLMs can be used to expand/explain/specify concise and ambiguous queries, thereby disambiguating the user information need. 
This bridges the gap between the latent user intent in the original queries and the retriever's representation of the user information need. 
The approach query2doc~\cite{wang2023query2doc}, generates pseudo-documents by few-shot prompting LLMs and concatenates them with the original query to form a new \textit{expanded query}.
However, there are two major limitations to using LLMs for generating plausible query re-writes.

% Researchers have explored the possibility of generating pseudo documents \cite{wang2023query2doc} for improving performance of dense retrieval. This approach is similar to pseudo-relevance feedback mechanism where instead of relying top-ranked documents for feedback, a transformer based generative model generates documents given the query as context. 

\mpara{The problem of misaligned query and documents.}
Many queries have multiple aspects, senses, and intent granularities.
When using LLM-based query expansions, the expanded query results in choosing one aspect among the many possible aspects.
This entails in a problem that the chosen intent by the expansion might not align with the ground-truth intent.
% The generated pseudo-document may contain multiple intents and may not aid in disambiguation of the user information need. 
This lack of alignment between the actual intent and the inferred intent is problematic during training.
Specifically, such intent non-alignment results in the case where the relevant document is forced to be matched with an erroneously expanded query.
% The approach also assumes the generated pseudo-documents have lexical overlap with ground truth documents which may not hold true for some scenarios. 
For instance, in Table~\ref{tab:intro_anecdotal}, we observe that for the query \texttt{define sri}, \qd{} drifts away from the intent and generates a query related to \textit{Stanford Research Institute }instead of denoting the Sanskrit word. In a similar manner, for the query \texttt{HS worms} the generated output reflects multiple intents. 
\begin{table*}
\begin{tabular}{lp{.82\textwidth}}
\toprule
    \textbf{Method}     & \textbf{Query Text} \\
\midrule
\textbf{Original query}  & \textbf{define sri}\\
\qd{}  & SRI stands for "Stanford Research Institute," which is a nonprofit research organization that was founded in 1946 as part of Stanford University. Over the years, SRI has conducted groundbreaking research in a wide range [\dots] \\
\vinci{}  & Sri is a Sanskrit word meaning “holy” or “auspicious.” It is often used in Hinduism as an honorific title for a deity or spiritual teacher. It is also used in Buddhist and Jain contexts, and is an important concept in South Asian culture. \\
\chatgpt{}  & Find the definition and meaning of the acronym SRI (Socially Responsible Investing).\\
\car (\vinci{})  & What is the meaning of Sanskrit word "Sri" and how is it used as a title of veneration? \\
\car (\chatgpt{})  & What is the meaning of the Sanskrit word "Sri" and how is it used in Indian subcontinent?\\
\car (\vinci{})(doc)  & What is the origin and meaning of the term "Sri"? \\
\car (\chatgpt{})(doc)  & What is the origin of the word "Sri" used in India and other South Asian countries?\\
\midrule
\midrule

\textbf{Original query}  & \textbf{hs worms}\\
\qd{}  & HS Worms can refer to several different things depending on the context. If you are referring to HS Worms as an educational institution, then it stands for Hochschule Worms which is a university of applied sciences or it can refer to heartworms [\dots] \\
\vinci{}  & Find information about heartworms in dogs\\
\chatgpt{}  & Find information about the symptoms, causes, and treatment options for heartworm disease in dogs (also known as HS worms).\\
\car (\vinci{})  & What are the courses offered by Hochschule Worms?\\
\car (\chatgpt{})  & What are the programs offered by Hochschule Worms?\\
\car (\vinci{})(doc)  & What is the country rank and world rank of HS Worms University?\\
\car (\chatgpt{})(doc)  & What is Hochschule Worms (HS Worms) and where is it located?\\
\midrule
\midrule

\textbf{Original query}  & \textbf{ephesians 6:5-9}\\
\qd{}  & Ephesians 6:5-9 is a passage from the New Testament that speaks to the relationship between slaves and masters in ancient times. It begins by instructing slaves to obey their earthly masters with respect [\dots] \\
\vinci{}  & Find information about Ephesians 6:5-9.\\
\chatgpt{}  & What is the meaning and interpretation of Ephesians 6:5-9 in the Bible?\\
\car (\vinci{})  & 7 With good will render service, as to the Lord, and not to men, 8 knowing that whatever good thing each one does, this he will receive back from the Lord\\
\car (\chatgpt{})  & How does the passage in Ephesians 6:5-9 instruct slaves to behave towards their masters?\\
\car (\vinci{})(doc)  & too have a Master in heaven. What does Ephesians 6:5-9 teach about the relationship between slaves and masters? \\
\car (\chatgpt{})(doc)  & What is the meaning of Ephesians 6:5-9 in relation to submission and slavery?\\
\midrule
\midrule

\textbf{Original query}   & \textbf{price nrz}\\
\qd{}  & Price NRZ stands for "Non-Return-to-Zero" and refers to a digital signal encoding technique used in telecommunications. In this technique, the voltage level of the signal remains constant during each bit interval[\dots] \\
\vinci{}  & Find information on the Pricing of Non-Recourse Z-Bonds.\\
\chatgpt{}  & Find information on the pricing strategy for non-return-to-zero (NRZ) encoding.\\
\car (\vinci{})  & What is the pricing of New Residential Investment Corp. common stock on Jan 30, 2017 7:16 PM EST?\\
\car (\chatgpt{})  & What is the public offering price of New Residential Investment Corp.'s common stock?\\
\car (\vinci{})(doc)  & announced today that it has entered into definitive agreements to acquire Shellpoint Partners, LLC ("Shellpoint"), a leading mortgage servicer and originator.hat is the price of New Residential Investment Corp (NRZ) stock \\
\car (\chatgpt{})(doc)  & What is the current share price and financial information of New Residential Investment Corp (NRZ)?\\
\midrule

\end{tabular}
\caption{Comparing \textbf{Original} ambiguous queries with their rewrites using \qd{}, \vinci{}, \chatgpt{} approach with in-context and context aware rewriter(\car{}) techniques. Approach with \textbf{(doc)} are rewrites using \ms{} document corpus and the rest from \ms{} passage corpus.}
\label{tab:intro_anecdotal}
\vspace{-0.25cm}
\end{table*}

\mpara{The problem of efficient inference.}
Secondly, a serious limitation (as acknowledged  in~\cite{wang2023query2doc}) is the computational overhead from both the retrieval and re-ranking phases.
Expanded query terms in the re-written queries increase index lookups.
More acutely, query expansions use tokenwise auto-regressive decoding during inference. 
However, current infrastructures just cannot support efficient LLM-based inference during query processing.
Therefore, to stay within sub-second ranking requirements without using prohibitively expensive compute, the choice of a LLM during query processing should be avoided.

\mpara{Contextualized query rewriting.}
We make two simple yet important design decisions to overcome the problem of misalignment and efficiency.
We first generate query rewrites by additionally providing the relevant document as context during training. 
Consequently, the generated query rewrite is fully aligned with the context improving training of text rankers.
Secondly, and unlike existing works, we fully avoid any query rewriting using LLMs \textbf{during inference}. 
In other words, we assume that training a ranker to match LLM-generated queries with relevant documents results in learning a generalized ranking model.

Methodologically, we propose an LLM-based context-aware query rewriting framework \car{}, that replaces the traditional query rewriter with a LLM for rewriting ambiguous queries. 
We employ context-aware prompting to test the effectiveness of LLMs as a disambiguation engine, using the ambiguous query and the relevant document as a context in the prompts. We also include a context selection mechanism to select relevant sections when a context is long and spans multiple topics to handle \textit{topic drift}.
The outputs of our query rewriter are used to fine-tune a ranking model to transfer the knowledge of user information needs to the ranking model. At \textbf{inference time}, the ranker equipped with knowledge of query disambiguation mechanism improves the document ranking performance for subsequent ambiguous queries without the rewriter component. The proposed framework can be used with any off-the-shelf ranker. 
% Our framework is different from query2doc \cite{wang2023query2doc} in various aspects. \textit{Firstly}, we perform \textbf{query reformulation} by generating short expansions grounded on relevant context and not large pseudo documents of ambiguous queries, limiting the possibility of topic drift. \textit{Secondly}, rewriter disambiguates queries only for fine-tuning the ranker and not during inference time, unlike query2doc. 
During inference, the ranking model fine-tuned on re-written queries yields much better ranking performance in comparison to the original queries. 
Our approach \textit{obviates the need for the LLM prompting during inference} because we consider efficiency and latency requirements during inference \textit{as a non-negotiable constraint}. 
From Table \ref{tab:intro_anecdotal} we observe that the proposed context-aware query reformulation approach \car{}, can generate concise rewrites that reflect the intent when compared to \qd{}. We also observe that it performs better than pure in-context learning based approaches, as seen from Table \ref{tab:intro_anecdotal} and also through our empirical analysis of ranking results.

We perform extensive experiments using LLMs of different parameter scales and demonstrate that the proposed approach results in better ranking performance. 
Specifically, we find that fine-tuning a ranker using re-written queries offers a significant improvement of up to \textbf{33\%} on the passage ranking task and up to \textbf{28\%} on the document ranking task when compared to the baseline performance of using original queries.
% The primary aim of the proposed approach is to disambiguate user intent of the under-specified queries through query reformulation for fine-tuning the ranker. 

\subsection{Research Questions}

We address the following research questions:

\mpara{RQ1}: Can we employ LLMs to generate fluent natural language rewrites of original queries from ambiguous queries?

\mpara{RQ2}: How effective is a ranker fine-tuned on rewritten queries for downstream document ranking task?
%How do they compare with ranker fine-tuned on classical queries in terms of performance?

%\blue{\mpara{RQ3}: How to deal with topic drift in long documents?}

%\blue{\mpara{RQ4}: How do we improve performance on hard queries?}

Towards answering these research questions, we do extensive experiments on trec web 2012 using LLM models (Section~\ref{models}) for evaluating quality of rewrites and on \trecdl{}-19 and \trecdl{}-20 passage and document dataset for re-ranking. 

\subsection{Contributions}

In summary, here is a list of our contributions: 
\begin{enumerate}
    \item We propose a Context Aware Rewriter (\car{}) for query reformulation, based on context-aware prompting of Large Language Models (LLMs) which generate natural language rewrites for \textit{ambiguous queries}. The natural language rewrites are used to fine-tune the ranker for encoding the ability to disambiguate the user information need.
    
    \item We experiment with LLMs of different parameter scales and with different pre-training objectives to demonstrate the difference in quality of rewrites.
    
    \item We show that the ranking model fine-tuned on generated rewrites delivers substantial performance improvement on the ambiguous user queries at inference time.

\end{enumerate}

\section{Related Work}
\label{sec:relatedwork}
\begin{figure*}
    \centering
    \includegraphics [width=0.8\textwidth]{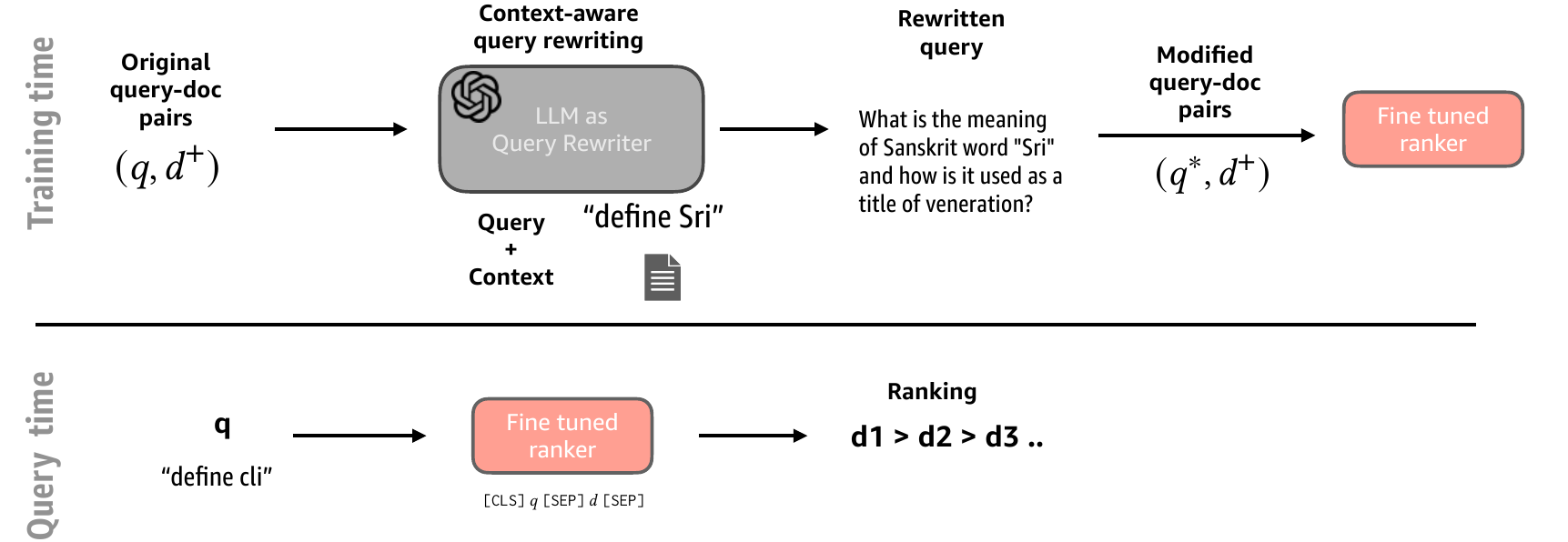}
    \caption{Overview of the proposed CAR framework of contextual re-writing queries }
    \label{system}
\end{figure*}
In this section, we discuss two aspects of related work, namely query expansion and generative capabilities of Large Language Models. We further discuss how LLMs are leveraged in addressing the vocabulary mismatch problem in information retrieval.
\subsection{Query rewriting}
One of the central problems in IR is bridging the lexical gap between the user specified query and the documents. This can also be seen as reconciling the difference between user intent and the intent of the system (\textit{machine intent}). The common approaches of query reformulation involve query expansion, synonym substitution and paraphrasing. 

Query expansion approaches have been adopted to bridge this gap. These approaches typically involve addition of terms to the original query based on relevance feedback \cite{relevance_lm, rocchio71relevance}. When user relevance feedback is unavailable, pseudo-relevance feedback mechanism is applied \cite{lv2009comparative, ir_query_exp}. Here, the top-ranked results to the original query are used to expand the query with additional terms. However, the performance of the rewritten query is severely limited by the quality of the top-ranked results \cite{prf_effectiveness} and is rarely used in dense retrieval \cite{ms_marco}. Alternatively, researchers have proposed rephrasing the original queries to tackle the ``lexical chasm" problem. In the work \cite{zukerman-raskutti-2002-lexical}, the authors rephrase the user specified query iteratively by mining similar phrases from WordNet. Alternatively, researchers have also explored substituting terms in the input query with synonymous terms based on user query logs \cite{query_substitution}. However, these approaches are term based and expand queries based on static sources, making it difficult to adapt to changes in the retrieval system. 

The generative approaches in query rewriting involve generating paraphrases of the original queries. In the work \cite{riezler-etal-2007-statistical}, the authors employ statistical approaches to rephrase terms of the query and add the equivalent terms to the original query. however, this could lead to ill-formed queries, as the term level paraphrasing does not consider the surrounding context. It also heavily relies on user feedback to select the best paraphrased version which is cumbersome. More promising approaches in query expansion involve paraphrasing queries using a generative model \cite{zerveas2020brown} at once instead of phrasal level rewrites. For instance, the query ``average tesla cost" is rephrased to ``what is the cost of the new tesla". Other generative approaches use chain-of-thought and relevance feedback ~\cite{jagerman2023query} to expand query terms or use pseudo relevance feedback along with prompting or fine-tuning ~\cite{wang2023generative} for query reformulation. However, this approach just provides alternative formulations of query leveraging equivalent queries data from MS MARCO dataset \cite{ms_marco}. They do not focus on disambiguation of user intent in the ambiguous queries. They also do not consider performance prediction on downstream retrieval and is also susceptible to exposure bias. To enhance the performance on downstream retrieval, DRQR \cite{wang2020deep} leverages deep reinforcement learning to train recurrent models by incorporating query performance predictors as reward signals. However, the authors mention that the proposed approach does not improve downstream retrieval by a significant margin.

More recently, natural language question generation approaches have been adopted for query reformulation. \cite{rao2018learning} introduced a model to generate  
clarifying questions to compensate for the missing information. They used a reinforcement learning model to maximize a utility function based on the value added by the potential response to the question. \cite{trienes2019identifying}, on the other hand, focused on identifying unclear posts in a community question answering setting that require further clarification. Query rewriting approaches have been of huge interest in the e-commerce domain and mostly require relevance feedback from the user for personalization \cite{query_expansion_e_commerce, taobao_search, zuo2022contextaware}. While these methods are domain-specific, \cite{zamani2020generating} propose a template-based weak supervised learning approach for open domain IR.

\textit{Document expansion} approaches have also been proposed to tackle the vocabulary mismatch issue in IR. Doc2query \cite{doc2query} and docTTTTquery \cite{docT5query}, generates pseudo queries from document context using a seq2seq model and appends them to documents to enhance the ranking results. Works like InPars~\cite{bonifacio2022inpars} and PromptAgator~\cite{dai2022promptagator} focus on using large generative models to generate queries from sampled documents to increase training data for dense retrieval tasks.  However, these approaches are prone to hallucination and may drift away from the original intent of the query \cite{gospodinov2023doc2query}. In this paper, we take a step in the direction of rewriting ambiguous queries by incorporating relevant context to transfer the knowledge of the disambiguation mechanism to a ranker for improved re-ranker performance. Our approach also departs from the traditional approaches by employing a monolithic LLM for query rewriting than multiple components that could lead to error propagation.

\subsection{Large Language Models}
Recent advances in generative modeling have led to large language models that apply to a wide range of tasks \cite{brown2020language, wei2021finetuned, alayrac:2022:flamingo}. These models trained in a  self-supervised fashion demonstrate emergent capabilities where they can extrapolate to new tasks with few demonstration samples \cite{gpt3incontext, incontext_survey, wei2023larger}. These advances have also led to researchers rethinking parts of retrieval systems \cite{zhu2023large}. Generative retrieval is one such emerging paradigm \cite{Metzler_2021, bevilacqua2022autoregressive, tay2022:differentiable:index,lee2022generative} where neural language models are used as indices for retrieval. The model employs conditional decoding approaches to generate document identifiers that directly map to ground truth documents. This is accomplished by training the language models on relevant data.

More recently, researchers have tested the effectiveness of LLMs for document expansion \cite{hyde, wang2023query2doc} in a zero-shot and few-shot setting for dense retrieval. However, these approaches have serious limitations at inference time in terms of efficiency due to the autoregressive decoding of LLMs. These approaches are also prone to hallucination due to the generation of long context for ranking.  Query rewriting using LLMs has also found interest in QA tasks with the new rewrite-retrieve-read paradigm \cite{ma2023query}. In this work, the authors  fine-tune a small LLM for query rewriting by employing reward signals from the reader. However, this approach is specific to QA tasks and the retriever is a frozen model, relying on the reader for fine-tuning the rewriter. The approach also requires the expensive query rewriting mechanism during inference.Researchers have also explored task-aware tuning of dense retrievers \cite{asai2022taskaware} by appending queries with task-specific instructions as input to the dense encoder. The instructions equip the retriever with the ability to decipher user's information needs. However, the instructions require manual annotation and are limited to specific tasks. In this paper, we make the ranking model aware of disambiguation mechanism of under-specified queries by fine-tuning it on rewritten queries generated by a LLM. 
%\todo {Add more recent references from https://arxiv.org/pdf/2308.07107.pdf}
\begin{figure}
   \centering
    \includegraphics[width=0.7\linewidth]{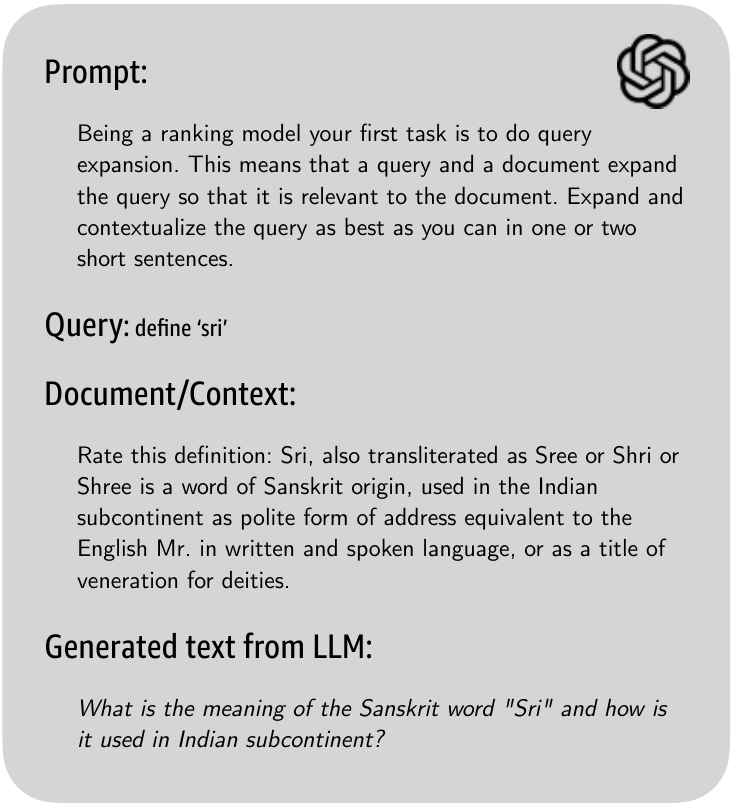}
    \caption{An example of prompting LLM for query rewriting using \car{}.}
   \label{prompting}
\end{figure}
% \begin{table}[]
%     \centering
%     \begin{tabular}{p{.50\textwidth}}
%        \toprule
%        \textbf{{\Large Prompt:}} \\
%        Being a ranking model your first task is to do query expansion. This means given a query and a document expand the query such that it is relevant to the document. Expand and contextualize query as best as you can in one or two short sentences. \\
%        \midrule
%         \textbf{\textcolor{red}{ Query:}} define `sri' \\
        
%      \textbf{\textcolor{red}{ Document:}} Rate this definition: Sri, also transliterated as Sree or Shri or Shree is a word of Sanskrit origin, used in the Indian subcontinent as polite form of address equivalent to the English Mr. in written and spoken language, or as a title of veneration for deities. \\    
%      \midrule
% \textbf{\textcolor{red}{ \Large Generated text from LLM:}} \\
%      What is the meaning of the Sanskrit word ""Sri"" and how is it used in Indian subcontinent? \\
%      \midrule
%     \end{tabular}
%     \caption{An example of context-aware prompting of LLM for query reformulation}
%     \label{tab:prompting}
% \end{table}
\section{Method}
In this section, we describe the proposed Context Aware Rewriter (\car{}) framework for ambiguous query reformulation. Figure \ref{system} shows the working of the proposed framework during the training and inference stages. During training, the framework is composed of a query reformulation phase and a document ranking phase. In the query reformulation phase, we employ a context-aware prompting of a LLM (Section \ref{rewriter:method}). During inference, the ranker which was fine-tuned on disambiguated queries is directly employed to rank documents for new queries without rewriting them, as shown in Figure \ref{system}.

\subsection{Query Rewriter}
\label{rewriter:method}
Algorithm~\ref{alg:augmentation} details the query reformulation process. Given an ambiguous query $q$, our goal is to generate a rewrite $q^*$ that helps disambiguate the intent of the original query. We generate query rewrites by conditioning the LLM ($RE$) through few-shot prompting on the relevant document for the query $d^+$ to avoid topic drift. An example of the few-shot prompting employed is shown in Figure~\ref{prompting}.
\[q^* = RE(q,d^+)\]
We first instruct the LLM (Figure~\ref{prompting}) to perform the task of query reformulation in the context of the given document. We then provide an ambiguous query ($q$) concatenated with the corresponding relevant document as a part of the prompt. This \textit{context-aware prompting} of LLMs results in better rewrites without topic drifts, as the LLM is conditioned on intents conveyed in the document. We also introduce a constraint on the maximum length of the output sequence generated. This constraint coupled with grounding of the generation on the relevant document context through prompting to prevent hallucination and topic drift.

Note that the intent conveyed for the rewrite can be controlled by using a different document context in the prompt. We experiment with different prompting approaches by varying the document context. We discuss the different variations of few-shot prompting and the corresponding results in Section~\ref{models} and Section~\ref{results} respectively.
\subsubsection{Hyperparameters}
We use a temperature of 0.5 to balance exploration and deterministic generation. We set presence penalty and frequency penalty to 0.6 and 0.8 respectively to minimize redundancy in generated reformulations of the original query.

The resulting rewritten query is used as an input to fine-tune the ranker. Note that we only rely on LLMs to rewrite ambiguous queries for fine-tuning the ranker. During inference, the ranker directly ranks relevant documents for the new queries without the LLM based rewriting component. 

\subsection{Ranking Phase}
\label{ranker:method}
Our goal is to train a model for \textit{re-ranking} documents on the query rewrites for under-specified queries. Given a query-document pair $(q^*,d)$ as input, the ranking models output a relevance score. This score can then be used to rank documents based on their relevance to the given query.

\begin{algorithm}[t]
    \DontPrintSemicolon
    \SetKwFunction{Rewrite}{rewrite}
                \tcp{Training Phase}

    \SetKw{In}{in}
    \KwIn{training batch $D$}
    \KwOut{training batch with rewritten queries $D'$}
    $D' \leftarrow$ empty list\;

    \ForEach{$(q, d^+)$ \In $D$}{
        \tcp{create rewritten query}
        $q^* \leftarrow RE(q,d^+)$\;
        append $(q^*, d^+)$ to $D'$\;
        Fine-tune ranker on $D'$
    }
    \Return{$D'$} \\
                \tcp{Inference Phase}
            \SetKw{In}{in}
    \KwIn{test set $D_t$}
    \KwOut{Ranked documents}
               \ForEach{$(q_t,d)$ \In $D_t$}{
               $\hat{y} \gets (q_t,d)$
} 
    \caption{Query Rewriting, Fine-tuning and Ranking}
    \label{alg:augmentation}
    
\end{algorithm}

Formally, the training set comprises pairs ${q^*_i, d_i }$, where $q^*_i$ is a rewritten (disambiguated) query using an LLM and $d_i$ is a relevant or irrelevant document to the query. The aim is to fine-tune a ranker $R$ that predicts a relevance score $\hat{y} \in [0; 1]$ given a reformulated query $q^*$ and a document $d$:
\begin{equation}
R: (q^*, d) \mapsto \hat{y}
\end{equation}

The fine-tuned ranking model ($R$) can be employed to re-rank a set of documents obtained from a first-stage lightweight frequency-based retriever. Recent studies have indicated that pre-trained language models which jointly model queries and documents  tasks~\cite{nogueira_prr_2019,dai_sigir_2019, rudra2020distant} have demonstrated significant performance on ranking tasks. In this work, we employ the \bert{}\cite{devlin_bert_2018} model for ranking. The input to the ranker is of format:
\begin{equation}
    \texttt{[CLS]}\ q\ \texttt{[SEP]}\ d\ \texttt{[SEP]}.
\end{equation}
 We employ the pointwise loss to train the ranker. Let us assume of a mini batch of $N$ training examples $\left\{x_i, y_i\right\}_{i=1, ..., N}$. The ranking task is cast as a binary classification problem, where each training instance $x_i = (q^*_i, d_i)$ is a query-document pair and $y_i \in {0, 1}$ is a relevance label. The predicted score of $x_i$ is denoted as $\hat{y}_i$. The cross-entropy loss function is defined as follows:
\begin{equation}
    \mathcal{L}_{\mathtt{Point}} = -\frac{1}{N} \sum_{i=1}^{N} \left( y_{i} \cdot \log \hat{y}_{i} + (1 - y_{i}) \cdot \log (1-\hat{y}_{i}) \right)
\end{equation}

Note that we only fine-tune the ranker on query reformulations of ambiguous or under-specified queries. At inference time, the ranker equipped with the knowledge of user information needs is directly deployed to rank relevant documents for new queries as shown in \textbf{Algorithm \ref{alg:augmentation}}.

\subsubsection{Passage selector for Documents}: 
\label{ranker:attention_linear}
The phenomenon of \textit{topic drift} has been identified in the context of utilizing the \car{} (Conversational Response Ranking) approach with documents. \textbf{Topic drift} is characterized by the divergence of a rewritten query's specificity when applying the \car{} approach to documents. This occurs due to documents encompassing multiple topics, potentially relating to one or more aspects of the query. To mitigate this occurrence of \textit{topic drift}, we employ supervised passage selection techniques (\attention{}, \linear{}) as proposed in ~\cite{leonhardt2021learnt}. These techniques aid in selecting the most relevant passage from a document, aligning it more closely with a given query.

\begin{itemize}

\item  \textbf{Linear Selection}: This method involves the transformation of both the query and the passage into their average embedding representations. These representations are then processed through a singular feed-forward layer, followed by a calculation of their similarity through dot product operation. The sentence score, denoted as $s_{ij}$ with respect to the query $q$, is computed as follows:

\begin{equation}
    \operatorname{score_{Lin}}(q, s_{ij}) = \langle \operatorname{Enc}(q), \operatorname{Enc}(s_{ij}) \rangle,
\end{equation}
where $\langle \cdot, \cdot \rangle$ is the dot product.
    \item \textbf{Attention selection} : The Attention-based selector operates by deriving passage-level representations through the utilization of the \textsc{QA-LSTM} model \cite{tan2016improved}. Initially, both the query and the document undergo contextualization by being subjected to shared bi-directional Long Short-Term Memory (LSTM) processing of their token embeddings. Subsequently, the query representation $\hat{q}$ is derived through the application of element-wise max-pooling over these contextualized embeddings.
\begin{align}
    \hat{q} &= \operatorname{Max-Pool}(\operatorname{Bi-LSTM}(q)), \\
    d^\text{LSTM} &= \operatorname{Bi-LSTM}(d).
\end{align}

For each hidden representation $d^\text{LSTM}_i$, attention to the query is computed as
\begin{align}
    m_i &= W_1 h_1 + W_2 \hat{q}, \\
    h_i &= d^\text{LSTM}_i \exp \left( W_3 \tanh \left( m_i \right) \right),
\end{align}
where $W_1$, $W_2$ and $W_3$ are trainable parameters. For $s_{ij}$, let $h_{ij}$ denote the corresponding attention outputs. The sentence representation is computed similarly to the query representation, i.e.,
\begin{equation}
    \hat{s}_{ij} = \operatorname{Max-Pool}(h_{ij}).
\end{equation}
The final score of a sentence is computed as the cosine similarity of its representation and the query representation:
\begin{equation}
    \operatorname{score_{Att}}(q, s_{ij}) = \operatorname{cos}(\hat{q}, \hat{s}_{ij}).
\end{equation}
\end{itemize}

\newcommand{\trec}{\textsc{Trec}}

% \section{Experimental Setup}
\section{Experimental Setup}
\label{experiments}

In this section, we describe the setup we used to answer the following research questions:

\mpara{RQ1}: Can we employ LLMs to generate fluent natural language rewrites of ambiguous and under-specified queries?

\mpara{RQ2}: How effective is a ranker fine-tuned on rewritten queries using LLMs for downstream document ranking task?
%How do they compare with the ranker fine-tuned on classical queries in terms of performance?

%\blue{\mpara{RQ3}: How to deal with topic drift in long documents?}

%\blue{\mpara{RQ4}: How do we improve performance on hard queries?}

Towards answering these research questions we employ the following datasets, rankers and training settings
%To evaluate our approach, we first fine-tuned or prompted a LLM-based re-writer using benchmark queries to test the quality of rewrites. We then used the re-written queries, in place of the original queries, to fine-tune a ranker. It is worth noting that we did not have any explicitly provided training data for fine-tuning our re-writers. Therefore, we relied on the small amount of topic description data available in the TREC datasets. More details about the datasets, models, and experimental settings are provided to answer the above research questions.

\subsection{Datasets}

Our goal is to have natural language expansions of the query for downstream ranking. Since there are no existing datasets explicitly tackling this problem, we curate queries and their descriptions from several sources, as detailed below.

\subsubsection{Datasets for checking effectiveness of the LLM-based re-writer.}
To tackle \textbf{RQ1}, we collect several datasets from \trec{} web track \cite{clarkeoverview}. Particularly, we collect topics and corresponding descriptions from \trec{} web tracks from 2009 to 2011 to be used as dataset for fine-tuning the generative model.  
We sourced the \texttt{(topic, topic description)} pairs to train a generative model.
We obtain 1143 samples for training, 126 samples for validation. 
Finally, we use 103 samples of the \trec{} Web 2012 topics, subtopics, and their descriptions for testing the rewriter. 

\subsubsection{Google People Also Asked (PAA) Data}
\label{subsubsec:googlepaa}
Since the TREC topics/queries do not disambiguate the multiple intents a query could have, we propose to collect questions from the Google \textit{people also ask (PAA)} section as external knowledge for each query. These texts from Google PAA could serve as expanded versions of the ambiguous query that convey different intents. For instance, for a topic \textit{403b}, the proposed pipeline provides different questions like \texttt{``What are the withdrawal limitations for a 403b retirement plan?''} and \texttt{``What is the difference between a 401k and 403b?''}. We augment the query topics in the dataset with the corresponding Google PAA questions.
We fine-tune the rewriters like \bart{} and \gpt{} with topics concatenated with the corresponding Google PAA question. However, on the test set (TREC web 2012) we leverage only the topics as access to an external knowledge base cannot be assumed during real-world deployments.
% We obtain several variants of rewriters using the above two datasets as shown in Table \ref{tab:rewriter}.

%\subsubsection{Datasets for Ranking Experiments}:
%We employ ambiguous queries from the MS-MARCO dataset for the downstream ranking tasks. These experiments conducted directly help answer RQ2.

%For testing out the hypothesis that fine-tuning the retriever with rewritten queries helps improve downstream retrieval we select several ambiguous queries from MS-MARCO. We select about 1567  ambiguous and under-specified queries from the MS-MARCO collection. Here only the retriever is fine-tuned, and the query rewrites are obtained from the exisitng rewriters.
\subsubsection{Datasets for Ranking Experiments}:

\textbf{\ms{} (Passage dataset)}: We consider the passage dataset from the TREC Deep Learning track (2019)~\cite{nguyen:2016:msmarco}. We evaluate our model on \trecdl{}-19 and \trecdl{}-20 passage dataset, each containing 200 queries. For training, we select 1200 ambiguous/under-specified queries from \trecdl{}-19 passage training dataset, then we rewrite these query using our query rewrite models (Subsection \ref{models}). The ambiguous queries are selected based on heuristics like query length, acronyms and entities with varied semantics. Each of these rewrites correspond to one dataset. So we end up with 13 rewrite training dataset and one baseline dataset (original queries).

\textbf{\ms{} (Document dataset)}: Similar to above passage dataset, we evaluate on \trecdl{}-19  and \trecdl{}-20 document collection. For training we select 1200 ambiguous/under-specified queries from \trecdl{}-19 document training dataset and use exactly the same method as above to create training datasets using different approaches. 

%\textbf{\ms{} (\mshard{} dataset)}
%\textbf{\beir{}}: The BEIR (Benchmarking and Evaluating IR) dataset is a large-scale benchmark dataset for information retrieval (IR) research. The dataset includes several retrieval tasks, including ad-hoc retrieval, sentence retrieval, and question answering. 
%\todo{} : rewrite beir dataset
\begin{table}[]
    \centering
    \begin{tabular}{|l|c|}
    \hline
        \textbf{Model} & \textbf{\# parameters}  \\\hline
         \texttt{text-ada-001} (\ada{}) &  350M \\
          \texttt{text-babbage-001} (\babbage{})  & 3B\\
          \texttt{text-curie-001 }(\curie{}) & 13B \\
         \texttt{text-davinci-002} (\davinci{}) & 175B \\
         \texttt{text-davinci-003 }(\vinci{})  & 175B \\
         \texttt{gpt-3.5-turbo }(\chatgpt{})  & 154B \\

  \hline
    \end{tabular}
    \caption{LLM (GPT3 models) of different parameter scales}
    \label{tab:gpt3_scales}
\end{table}
\subsection{Rewriter Models}
\label{models}

We employ several generative models as the backbone for query rewriting to compare and contrast the quality of query rewrites. We employ two settings where the rewrites are smaller language models fine-tuned on TREC web dataset for generating rewrites or are LLMs which are prompted to generate rewrites.

\paragraph{\textbf{\gpt{}}} We fine-tune \gpt{} a transformer decoder based generative model with 117M parameters and 12 layers to generate topic descriptions from topics on the TREC web dataset. We use a learning rate of 3.2e-5, weight decay of 0.01, batch size of 16 and train the model for 6 epochs.

\paragraph{\textbf{\bart{}}} We employ \bart{} (base version) a denoising autoencoder pre-trained using the objective for reconstructing the corrupted text. As a result, \bart{} is able to generate robust fluent natural language queries from under-specified and ambiguous queries. We fine-tune \bart{} to generate topic descriptions from given input topics on TREC web dataset. We use a learning rate of 2e-5, batch size of 16 and train the model for 8 epochs 

\paragraph{\textbf{\bart{} (topic+PAA)}} We also propose a variation \bart{} (topic+PAA) where we concatenate the topics with external knowledge in the form of related Google PAA questions (c.f. Section \ref{subsubsec:googlepaa}) for fine-tuning the \bart{} base model on the TREC web dataset collected. However, when generating rewrites for new ambiguous queries in the MS MARCO dataset, using the fine-tuned model, we provide only topics as input. We propose this approach to test if augmentation of Google PAA questions as context to topic names during fine-tuning aid in disambiguation of the user information need for smaller models. This variant also categorizes to context-aware rewriting class of approaches \car{} proposed in this work, where Google PAA acts as context. We use similar hyperparameters for \bart{} as discussed. This is slightly different from the proposed prompting based \car{} approaches and is proposed as an alternative to test the capabilities of smaller models.

\paragraph{\textbf{\vinci{} (prompting)}}  Since prompting LLMs like \gptt{} have proven to yield more stable outputs with less factual errors, we employ two variants of prompts to \vinci{}. We feed the prompts \texttt{``Generate short sentence expanding:''} or \texttt{``Generate short sentence question:''} followed by the topic name to \vinci{}. We call these two variants \vinci{} (prompt1) and \vinci{} (prompt2) respectively. We use a temperature of 0.5, max token length of 35 to generate short natural language rewrites of the original query. For frequency penalty and presence penalty, we use values of 0.8 and 0.6 to avoid redundancy in generated outputs.

\paragraph{\textbf{\vinci{} + in-context examples}} Since, plain prompting might result in topic drifts of generated text and the model might misinterpret the intended task, we also adopt in-context learning \cite{gpt3incontext}. In-context learning treats the LM as a black-box and instructs the model of the task without gradient descent through examples. We provide examples in form of 
$$ \mathtt{<query_1,desc_1>}, \mathtt{<query_2, desc_2>} \ldots \mathtt{query_{test}, [insert]}
$$ 
and instruct the model to fill the description in the placeholder provided. We use the same hyperparameters as discussed for \vinci{} (prompting).

\paragraph{\textbf{Other \gptt{} models}} We also test with models which are variants of \gptt{} of different parameter scales. These models include \ada{}, \davinci{}, \curie{} and \babbage{}. The models and the number of parameters are shown in Table \ref{tab:gpt3_scales}. We employ in-context learning by providing demonstration samples. The prompt is as follows: \\
\textit{Generate short sentence as expansion for the given test query like the following examples,}
$$ \mathtt{<input: query_1,output: desc_1>}, \ldots \mathtt{input:query_{test}, output:}$$
Note that the prompt is a bit different from \vinci{} as we observed that smaller models need detailed instructions for better generation capabilities. This maybe due to the difference in instruction fine-tuning approaches. We employ the same hyperparameters, queries and their descriptions as discussed earlier.

\paragraph{\textbf{\chatgpt{} + in-context examples}}

We also test the in-context learning capabilities of gpt-3.5-turbo (\chatgpt{}) for query reformulation. We employ the same prompt as \vinci{}. We only prepend the task instruction to set the role of the system. The task description is :"\textit{You are a system that gives an expansion for queries, expanding abbreviations and acronyms when applicable. Some examples are}" followed by the demonstration samples as discussed earlier. We employ the same values for hyperparameters as discussed for other LLM prompting approaches.

\textbf{\qd{}}: We also use the recently proposed document expansion approach, \qd{} which generates pseudo documents to aid in document ranking, as a baseline. We use the gpt-3.5-turbo model with max tokens of 128 for generation and follow the original hyperparameters used in the work \cite{wang2023query2doc} for reproducibility.

\subsection{Ranking Models}
\label{ranking_models}
 For experiments on \trecdl{} we use \bert{}-base~\cite{devlin_bert_2018}, a pre-trained contextual model based on the transformer architecture.
 In principle, one can use different transformer architectures but focus on \bert{} as a representative model in our experiments.
 We use the \emph{base} version with 12 encoder layers, 12 attention heads and 768-dimensional output representations. The input length is restricted to a maximum of 512 tokens. 
 We use cross-attention \bert{} architecture for ranking, sometimes also referred to as \textsc{MonoBert}. 
 The baseline model is trained on the original set of ambiguous queries using a pointwise ranking loss objective. The improved ranking models are also BERT base models but trained on re-written queries using LLMs such as \gptt{} variants, \bart{}, \chatgpt{}, etc. In the experiments, we refer to these improved ranking models after the LLM query re-writer model that used to generate the training data for them.

%\subsubsection{Hyperparameters}

\subsection{Metrics}

To evaluate the quality of re-writes we use the TREC topic descriptions as targets and compute ROUGE-L~\cite{rouge2004package} scores and \textsc{BERTScore}~\cite{bertscore,codebertscore,bertscore_app} which are commonly used automated evaluation metrics used in text several generation tasks. While ROUGE-L is an n-gram overlap metric, BERTScore correlates with human judgments by employing contextualized embeddings to compute word-level matches between the reference sentence and the generated sentence~\cite{bertscore}.

The metrics serve as a proxy to measure the ability of the generative models to expand the query to disambiguate various intents to aid in downstream retrieval tasks. They also measure if the queries are plausible.
To evaluate the ranking approach, we employ standard ranking metrics such as MRR and nDCG@10 and evaluate on \trecdl{}-19 and \trecdl{}-20 test sets for passage and document collection.

\begin{table}
\begin{tabular}{llll|r}
\hline
\multirow{2}{*}{\textbf{Method}} & \multicolumn{3}{c|}{\textbf{BERTScore}} & \multirow{2}{*}{\textbf{Rouge-L}} \\
\cmidrule(lr){2-4}
  & \textbf{P}     & \textbf{R}     & \textbf{F1}    &  \\
\hline

GPT2 (topic){$^\dagger$}                   & 0.457          & 0.447          & 0.451          & 7.47            \\
\bart{} (topic){$^\dagger$}        & 0.792          & 0.791          & 0.792          & 28.89           \\
\qd{}  & 0.672 & 0.749 & 0.708 & 6.76 \\
\ada{}  (in-context)         & 0.734          & 0.776          & 0.754            & 17.43            \\
\babbage{} (in-context)           & 0.804          & 0.802          & 0.802          & 32.16            \\
\curie{} (in-context)          & 0.812          & 0.800          & 0.805          & 34.25            \\
\davinci{} (in-context)           & 0.818         & 0.823          & 0.820        & 33.81            \\
\vinci{} (prompt1)            & 0.727          & 0.743          & 0.734          & 14.38          \\
\vinci{} (prompt2)            & 0.811          & 0.792          & 0.801          & 32.07           \\
\vinci{} (in-context)         & 0.817 & 0.818 & 0.817 & 34.68  \\
\chatgpt{}           & 0.768          & 0.812          & 0.789          & 25.77           \\
\hline
\multicolumn{5}{l}{\bf  \car{}} \\
\bart{} (topic+PAA){$^\dagger$}  & 0.777          & 0.783          & 0.779          & 25.85           \\
\vinci{} &\underline{0.821} & \underline{0.834} & \underline{0.827} & \underline{38.57}  \\
\chatgpt{} & \textbf{0.825} &  \textbf{0.847} &  \textbf{0.836} & \textbf{40.51}\\ \hline
\end{tabular}
\caption{BERTScore and ROUGEL scores for different rewrites on TREC web 2012. The best results for each dataset and each model is in \textbf{bold} and second is \underline{underlined}. {$^\dagger$}indicates that the model has been fine-tuned with topic or topic + PAA data.}
\label{tab:rewriter}
\end{table}
\begin{table*}
    \centering
    \begin{tabular}{lp{.08\textwidth}p{.08\textwidth}p{.08\textwidth}p{.08\textwidth}|lp{.08\textwidth}p{.08\textwidth}p{.08\textwidth}p{.08\textwidth}}
    %\begin{tabular}{lcccc|lcccc}
        \toprule
            & \multicolumn{4}{c}{\textbf{\trecdl{} Passage}} && \multicolumn{4}{c}{\textbf{\trecdl{} Document}}\\
             \midrule
            & \multicolumn{2}{c}{\textbf{\trecdl{}-19}} & \multicolumn{2}{c}{\textbf{\trecdl{}-20}} && \multicolumn{2}{c}{\textbf{\trecdl{}-19}} & \multicolumn{2}{c}{\textbf{\trecdl{}-20}}\\
            \cmidrule(lr){2-3}
            \cmidrule(lr){4-5}
            \cmidrule(lr){7-8}
            \cmidrule(lr){9-10}
            \textit{Ranking Models}
            &  $\text{RR}$ & $\text{nDCG}_\text{10}$ &  $\text{RR}$ & $\text{nDCG}_\text{10}$ &&  $\text{RR}$ & $\text{nDCG}_\text{10}$ &  $\text{RR}$ & $\text{nDCG}_\text{10}$ \\
            \midrule

%% RR and nDCG@10 scores
\multicolumn{5}{l}{\bf Baseline} \\
\bert{} Baseline &  0.653 & 0.381 &  0.523 & 0.288 && 	0.750 &	0.453 & 	0.735 &	0.393 \\	
\qd{}~\cite{wang2023query2doc} &  0.578\down{11.6}$^{\#}$ & 0.321\down{15.7} &  0.608\up{16.2} & 0.323\up{12.3} && 	\textbf{0.82\up{9.7}} &	0.467\up{3.1}$^{*}$ & 	0.749\up{2} &	0.415\up{5.6}$^{*}$ \\
\midrule
\multicolumn{5}{l}{\bf Query expansion} \\
\bart{} (topic)     &  0.500\down{23.5}$^{*}$ & 0.254\down{33.3}$^{*}$ &  0.380\down{27.3}$^{*}$ & 0.202\down{29.9}$^{*}$ && - & - & - & - \\
\ada{} &  0.634\down{2.9} & 0.370\down{2.9} &  0.479\down{8.4} & 0.278\down{3.6} && - & - & - & - \\
\babbage{} &  0.697\up{6.7} & 0.370\down{3.0} &  0.580\up{10.9} & 0.327\up{13.5} && - & - & - & - \\
\curie{} &  0.563\down{13.9} & 0.328\down{13.9} &  0.498\down{4.8} & 0.248\down{13.8} && - & - & - & - \\
\davinci{} &  0.795\up{21.7}$^{*}$ & 0.391\up{2.6} &  0.507\down{3.1} & 0.284\down{1.3} && - & - & - & - \\
 \vinci{} (prompt1) &  0.714\up{9.4} & 0.386\up{1.3} &  0.493\down{5.7} & 0.289\up{0.4} && - & - & - & - \\
\vinci{} (prompt2) &  0.717\up{9.9} & 0.397\up{4.1} &  0.574\up{9.7} & 0.338\up{17.3} && - & - & - & - \\
\vinci{}  &  0.766\up{17.3}$^{\#}$ & 0.411\up{7.9} &  0.552\up{5.6} & 0.308\up{7.1} && 	\underline{0.818\up{9.1}} &	0.490\up{8.2}$^{\#}$ & 	0.764\up{4.0} &	0.465\up{18.5}$^{\#}$ \\
\chatgpt{} &  0.689\up{5.5} & 0.396\up{4.0} &  \underline{0.623\up{19.2}}$^{\#}$ & 0.348\up{20.7}$^{\#}$ && - & - & - & - \\	
\midrule
\midrule
\multicolumn{5}{l}{\bf \car{}} \\
\bart{} (topic+PAA) &  0.745\up{14.2} & 0.391\up{2.6} &  0.614\up{17.4}$^{*}$ & \underline{0.363\up{26.0}} && 	0.762\up{1.6} &	0.504\up{11.3}$^{\#}$ & 	\underline{0.819\up{11.5}} &	\underline{0.485\up{23.4}}$^{\#}$ \\
\vinci{} & \textbf{0.798\up{22.2}}$^{*}$ & \underline{0.417\up{9.5}} &  0.533\up{2.0} & 0.316\up{9.9} &&	0.774\up{3.3} &	0.507\up{12.0}$^{*}$ & 	0.777\up{5.8} &	0.447\up{13.7}$^{*}$ \\
\chatgpt{} &  \underline{0.794\up{21.6}}$^{*}$ & \textbf{0.494\up{29.7}}$^{*}$ &  \textbf{0.653\up{24.8}}$^{*}$ & \textbf{0.383\up{33.1}}$^{*}$ && 	0.769\up{2.6} &	\underline{0.509\up{12.3}}$^{\#}$ & 	0.7826\up{6.5} &	0.430\up{9.4}$^{*}$ \\
\vinci{} (\attention{}) & - & - & - & - && 0.81\up{8.1} & \textbf{0.517\up{14.1}}$^{*}$ & 0.785\up{6.9} & 0.459\up{17}$^{\#}$ \\
\vinci{} (\linear{}) & - & - & - & - && 0.782\up{4.3} & 0.497\up{9.7}$^{*}$ & \textbf{0.848\up{15.5}} & \textbf{0.504\up{28.4}}$^{*}$ \\
\chatgpt{} (\attention{}) & - & - & - & - && 0.759\up{1.3} &	0.465\up{2.6} & 0.773\up{5.3} &	0.417\up{6.2}$^{*}$ \\
\chatgpt{} (\linear{}) & - & - & - & - && 0.778\up{3.8}	& \textbf{0.517\up{14.1}}$^{*}$ & 0.799\up{8.8} &	0.461\up{17.5}$^{*}$ \\

        \bottomrule        
    \end{tabular}

    \caption{Ranking performance of \bert{} model trained using query expansions from different query generative models on \trecdl{} '19 and \trecdl{} '20. We show the relative improvement of our approaches against a baseline (fine-tuned on original queries without query expansion) in parentheses. Statistically significant improvements at a level of $95\%$ and $90\%$ are indicated by $*$ and $\#$ respectively~\cite{paired_significance_test}. The best results for each dataset and each model is in \textbf{bold} and second is \underline{underlined}.}
    \vspace{-5mm}
    \label{tab:reranking-msmarco}
\end{table*}

\section{Results}
\label{results}
We begin by first answering if the re-writer component is able to generate plausible natural language expansions of queries. Then we conclude by analyzing the impact of the rewritten queries on the document ranking performance.

\subsection{Query Rewriter Evaluation}
%\textbf{How are the quality of rewrites ?}:
To answer \textbf{RQ1}, we evaluate the proposed rewriting approaches using metrics like BERTScore and ROUGEL.  The results are reported in Table~\ref{tab:rewriter}. We report  the mean values of precision, recall and F1 scores across all the test samples from TREC web 2012 to report the final scores. We observe that, \car{} variants, specifically, \vinci{} and \chatgpt{} approaches with context-aware prompting have the highest BERTScores and ROUGE-L scores. We attribute this performance to the \textbf{scale of the \gptt{}} language model, instruction fine-tuning and the \textbf{quality of context} provided that enable the rewriter to generate natural language queries. We observe that \car{} provides more fluent and relevant rewrites when compared to vanilla prompt based methods. The in-context learning based approaches also yield fluent rewrites. However, on analysis, we observe that they sometimes have topic drifts due to lack of relevant context needed to disambiguate the queries. We also observe that LLMs of smaller scales like \ada{}, \babbage{} and \curie{} generate rewrites which are factually inconsistent and drift away from the relevant topic. This is evident from the downstream document ranking performance(Table~\ref{tab:reranking-msmarco}) and the qualitative analysis of samples(Table~\ref{tab:intro_anecdotal}). This demonstrates that \car{} rewrites are consistent, fluent and help improve downstream ranking performance.

Among the fine-tuned approaches in Table \ref{tab:rewriter},  \bart{} (topic), \bart{} (topic + PAA) and \gpt{} (topic), we observe that \bart{} (topic) generates fluent query rewrites when compared to, \gpt{} (topic), as evident from BERTScore and ROUGEL scores. After performing a manual analysis of samples, and we observed that the query rewrites generated by \gpt{} were not relevant due to hallucination and in certain cases they were also grammatically incorrect. Therefore, we omit the \gpt{} (topic) model for the rest of the experiments. The \car{} (\bart{} (topic + PAA)) further improves the fluency of generated queries, but still falls short compared to \gptt{} variants and \chatgpt{} approaches, even though they are not fine-tuned.

Since the metrics proposed above are not true indicators of quality, we assess their quality through their impact on downstream ranking performance. 

\mpara{Insight 1}: LLMs can produce good rewrites for under-specified and ambiguous queries. Among our methods \car{} with context-aware few-shot prompting has the best rewrites. \\

\subsection{Ranking Evaluation}

To answer \textbf{RQ2} we first train rankers on data from \ms{} passage dataset (Section~\ref{ranking_models}) with rewrites from different LLMs (Section~\ref{models}), and evaluate on \trecdl{}-19 and \trecdl{}-20 test sets. Then we choose the best rewriter models and train document ranking models on \ms{} document dataset and evaluate on \trecdl{}-19 and \trecdl{}-20. All results are shown in Table~\ref{tab:reranking-msmarco}.
%To answer \textbf{RQ2} we fine-tuned a \bert{}-based ranker on the default queries(ambiguous) from \ms{} passage and document train set (refer Section \ref{experiments}) as a baseline. For the rest of our models, the queries are rewrites from models in Table~\ref{tab:rewriter}. We evaluate all models on \trecdl{}-19 and \trecdl{}-20 datasets. We do not rewrite queries from the test dataset. Table ~\ref{tab:reranking-msmarco} shows the performance of different rankers trained using ambiguous queries from \ms{} passage and document collection on \trecdl{}-19 and \trecdl{}-20. 

In case of passage dataset, CAR models perform better than their counterparts with \chatgpt{} \car{} showing improvements of \textbf{30\%} and \textbf{33\% }on nDCG@10 and \textbf{22\%} and \textbf{25\%} on MRR over baseline for \trecdl{}-19 and \trecdl{}-20 respectively. Apart from \car{}, \vinci{} in-context and \chatgpt{} in-context model outperform the baseline across test sets. 

The poor performance of models such as \ada{},\babbage{}, etc are evident from our observation that they are not able to disambiguate the query, but rather just rewrite it. For example, the query ``define sri'' is generally rewritten as ``Definition of sri'' while \vinci{} and \chatgpt{} are able to convey the intent better as shown in Table ~\ref{tab:intro_anecdotal}. Our \car{} models perform better than non \car{} models, proving that localized context is important along with global knowledge.
We also observe that \bart{} base model performs the worst among all models but variation of \bart{}, with context awareness \bart{}(topic+PAA) does better than baseline model in all evaluations. We posit that the external knowledge provided in the form of \textit{Google PAA} questions when fine-tuning \bart{}(topic+PAA) helps disambiguate the intent of the query. For instance, the query "403b" in TREC web track 2012 is ambiguous without context. However, one of the corresponding Google PAA questions, "What are the withdrawal limitations of a 430b retirement plan?" helps indicate that the query refers to a retirement plan. We posit that the \bart{} (topic+PAA) approach encodes this knowledge through fine-tuning. Hence, it has better document ranking performance compared to other approaches. This proves that context matters when it comes to rewriting ambiguous queries.

We also test our \car{} approach on \ms{} document collection. We consider the baselines and the best-performing approaches on the passage dataset to test on \ms{} document collection. Table ~\ref{tab:reranking-msmarco} shows the performance of different rankers trained using ambiguous queries from \ms{} document collection on \trecdl{}-19 and \trecdl{}-20. In the case of \trecdl{}-19 \chatgpt{} \car{} methods perform better than in-context models and for \trecdl{}-20 \bart{}(topic+PAA) performs better. \car{} models perform about \textbf{12\% } and \textbf{23\%} better than baseline model on nDCG@10. In the case of the passage dataset for a given query and a relevant passage, the context is specific, whereas this is not true in the case of document collection. There can be context drift in the case of a document, where a document can have generic contexts, which can result in LLM generating queries of generic intents using the \car{} approach. In Table ~\ref{tab:intro_anecdotal} we can see examples where in case of \car{} rewrites from different models are more specific compared to their document counterparts for the same queries. 

If we look at model performance using \ms{} passage collection (Table ~\ref{tab:reranking-msmarco}) and document collection (Table~\ref{tab:reranking-msmarco}) combined, we can see that \car{} approach performs better overall. This is due to the reason that the LLMs  employed to expand ambiguous/ill formed queries encode world knowledge, which when enriched with relevant passage/document context helps ground the generation. The context serves as grounding, which prevents hallucination and topic drift. In Table ~\ref{tab:intro_anecdotal} we can see how CAR model rewrites are more specific and factual compared to their non-CAR counterparts. 

\mpara{Insight 2}: The quality of context employed in few-shot prompting of LLMs is crucial when expanding ambiguous/ill formed queries. The proposed approach \car{}, is able to form concise and relevant query reformulations based on the given context.

\mpara{How do we deal with context drift for long documents?}

We previously discussed the phenomenon of \textit{topic drift} within the scope of document context. To mitigate this phenomenon, we employ techniques involving supervised  selection techniques, namely \attention{} and \linear{} selectors (Subsection~\ref{ranker:attention_linear}), as proposed in ~\cite{leonhardt2021learnt}. In both instances, the primary objective of the selector is to identify and select the most pertinent passage based on a given query.

In our methodology, given a query and a relevant document to the query, we segment the document into passages, each comprising approximately four sentences. Subsequently, we employ \attention{} and \linear{} selectors to identify the passage that bears the highest similarity to the given query within the document. This selected passage, along with the original query, is used to rewrite the query using LLM and then subsequently use it for model training.

The final four models detailed in Table~\ref{tab:reranking-msmarco} exemplify the application of \attention{} and \linear{} selectors on \vinci{} and \chatgpt{}, resulting in improved performance as compared to their \car{} counterparts. Just comparing \car{} models, we can clearly see \car{} \chatgpt{} (\linear{}) does 2\% and 8\% better than \car{} \chatgpt{} for $\text{nDCG}_\text{10}$ on \trecdl{} '19 and \trecdl{} '20 respectively. We cans see a similar trend in case of \vinci{} and \vinci{} (\attention{}) models. Empirical observations demonstrate that models employing the \car{} selector exhibit a performance enhancement ranging from 2\% to 15\% when compared to their \car{}-only counterparts. Hence, proving that if we reduce \textit{topic drift} for documents we can improve query rewriting quality, in turn improving downstream task performance. From Table ~\ref{tab:reranking-msmarco} we can concur that for query rewriting using documents \car{} based models perform the best, specifically \chatgpt{} (\linear{}) and \vinci{} (\attention{}) are the two best performing models overall.

\mpara{Insight 3}: Emphasizes that the utilization of \attention{} and \linear{} selectors to mitigate document topic drift not only enhances the quality of the rewritten query using the \car{} methodology, but also leads to an overall improvement in the ranking model's performance.

%\textbf{Is there a better way to rewrite query using LLM?} \\
%\textbf{Importance of doc length (number of context) to the quality of document context rewrite}: \\
%\textbf{Performance of other few-shot prompting approaches}:\\
\mpara{How does performance vary based on LLM of different parameter scales?}:
We experiment with models of different parameter scales (Table \ref{tab:gpt3_scales}). The results are shown in Table \ref{tab:reranking-msmarco}. We observe that as we scale up the rewriter LLM from 350 million parameters to 175 billion or 154 billion parameters, the performance improves significantly as observed during evaluation (Table \ref{tab:reranking-msmarco}). This is because they are able to characterize the world knowledge better from the instruction based training regime and the scale of the models \cite{gpt3_knowledge_base}. We observe that smaller models like \ada{}, \babbage{} and \curie{} generate factually inconsistent rewrites of the original query and rewrites with significant topic drift from the original query. 

\mpara{Insight 4}: We observe that when we scale up LLMs it results in better quality rewrites as they encode more world knowledge. This is  evident from the significant improvement in ranking performance.

%Answer the following too: 1. Why does \gptt{}(incontext) do better than all. 2. Why do \chatgpt{} do worse than \davinci{}. 3. Why does nDCG go down? 4. What does high MAP, RR but low NDCG tell us about the model?
%Expand observations: 1. GPT3 models are sensitive to prompts as seen from 3 different prompt models.

%\mpara{Insight 2}. Query re-writing helps boost ranking performance with \gptt{}(in-context) being the best ranking model.

%\mpara{How do we improve performance on hard queries(\mshard{} dataset)?}:

\subsection{Qualitative Analysis}
%We analyze examples of ambiguous queries and their corresponding rewrites obtained from the generative models. Some of these examples are in Table \ref{tab:intro_anecdotal}. We observe that the generated queries are close to actual intents in the \car{} (\chatgpt{} or \vinci{}) approach. For instance, for the query \textit{hs worms}, out many possible contexts such as the \textit{educational institution} or \textit{heartworms}. However, the corpus only contains documents pertaining to the educational institution. This renders the query and underlying intent ambiguous without being contextualized by the documents in the corpus. However, among the other rewriting approaches, \car{} is able to decipher the correct intent with fine granularity by also providing an expansion for the query. The advantage of the proposed framework is more evident from the fourth example \textit{urodeum function}. This query is also of ambiguous nature, as it could refer to any anatomy. However, we observe that \car{} with document context-aware prompting is able to decipher the correct intent. We attribute this to the large scale pre-training data and the ability of over-parameterized models to serve as knowledge bases \cite{gpt3_knowledge_base}.

We analyze examples of ambiguous queries and their corresponding rewrites obtained from the generative models. Some of these examples are in Table \ref{tab:intro_anecdotal}. We observe that the generated queries are close to actual intents in the \car{} (\chatgpt{} or \vinci{}) approach. For instance, for the query \textit{hs worms}, the corpus only contains documents pertaining to the educational institution. But it has multiple contexts, such as the \textit{educational institution} or \textit{heartworms}. This renders the query and underlying intent ambiguous without being contextualized by the documents in the corpus. However, among the other rewriting approaches, \car{} is able to decipher the correct intent with fine granularity by also providing an expansion for the query. The advantage of the proposed framework is more evident from the fourth example \textit{urodeum function}. This query is also of ambiguous nature, as it could refer to any anatomy. However, we observe that \car{} with document context-aware prompting is able to decipher the correct intent. We attribute this to the large scale pre-training data and the ability of over-parameterized models to serve as knowledge bases \cite{gpt3_knowledge_base}.

\mpara{Insight 5}: \car{} with context-aware prompting, is better at generating plausible rewrites with intents that are relevant to the search domain and improve downstream ranking.
\subsection{Limitations} In our current approach, we choose ambiguous queries based on heuristics like the query length and specific types of queries like acronyms, entities with multiple meanings based on context. Though these heuristics reflect the nature of ambiguous queries, a more principled approach would help identify and filter ambiguous queries.
\label{limitations}

%\input{algo.tex}
% \balance
\section{Conclusion}
\label{conclusion}
In this work, we propose a framework for redefining query rewriting approaches using context-aware prompting of LLMs for improving document ranking. Our framework reformulates ambiguous queries to interpretable natural language queries that disambiguate user information needs. Our experiments demonstrate that generating plausible rewrites which disambiguate user intent is possible, which further enhances downstream ranking performance. We posit that the joint training of the rewriter and ranker in our framework with feedback signals would yield better rewrites and ranking performance. We also propose several challenges associated with the development of the framework.
%We also propose several research directions, and challenges associated with the development of the framework.
%We will open source all our data, code, and prompts under:
%\texttt{\url{http://github.com/<MASKED>}} for reproducibility.    

% \clearpage
\balance
\newpage

\bibliographystyle{ACM-Reference-Format}
\bibliography{reference,interpretable-ir} 
% \input{appendix}
%\balancecolumns

\end{document}